\documentclass[aps,prb,twocolumn,showpacs]{revtex4-1}

\usepackage{graphicx}% Include figure files

\begin{document}

\title{\bf Fluxon Dynamics of a Long Josephson Junction \\
with Two-gap Superconductors}
\author{Ju H. Kim, Bal-Ram Ghimire and Hao-Yu Tsai}
\affiliation{
Department of Physics and Astrophysics, University of North Dakota,
Grand Forks, ND 58202-7129}

\begin{abstract}
 
We investigate the phase dynamics of a long Josephson junction (LJJ) with 
two-gap superconductors.  In this junction, two channels for tunneling between 
the adjacent superconductor (S) layers as well as one interband channel 
within each S layer are available for a Cooper pair.  Due to the interplay 
between the conventional and interband Josephson effects, the LJJ can exhibit 
unusual phase dynamics.  Accounting for excitation of a stable 2$\pi$-phase 
texture arising from the interband Josephson effect, we find that the critical
current between the S layers may become both spatially and temporally 
modulated.  The spatial critical current modulation behaves as either a 
potential well or barrier, depending on the symmetry of the superconducting
order parameter, and modifies the Josephson vortex trajectories.  We find that
these changes in phase dynamics result in emission of electromagnetic waves as
the Josephson vortex passes through the region of the 2$\pi$-phase texture.  We 
discuss the effects of this radiation emission on the current-voltage 
characteristics of the junction.    

\end{abstract}


\maketitle

\section{introduction}

Recently discovered\cite{Iron} iron-based superconductors have renewed much 
interest in the Josephson tunnel junctions involving multi-gap
superconductors.\cite{Suhl,Legg}  Both experimental and theoretical studies
indicate that several other superconductors, including $MgB_2$,\cite{MgB2, Xi} 
$NbSe_2$,\cite{NbSe2} heavy fermion $UPt_3$,\cite{UPt3} organic\cite{Organic} 
$(TMTSF)_2X$ and $\kappa-BEDT$, may have multiple pesudo-order parameters.  As
the appearance of a phase texture and unusual Abrikosov vortex 
properties\cite{Be} in two-gap superconductors are considered as manifestations
of the multi-component order parameter, the tunneling property of a Josephson junction with these superconductors may exhibit important differences from that
with one-gap superconductors.  This difference is attributed to the two tunneling channels for electrons in two-gap superconductor junctions due to the
presence of two condensates.  This internal freedom reflects the number of
electronic bands participating in superconductivity.  Consequently, dynamics of
the relative phase of the order parameters for the two-gap superconductor 
tunnel junction differ from that of the one-gap superconductor junction.

The presence of two tunneling channels indicates that there are two types of 
relative phase dynamics in a long Josephson junction (LJJ) with two-gap 
superconductors.   These phase dynamics may be understood in terms of the 
interplay between the interband and conventional (intraband) Josephson effects.
The interband Josephson effect describes tunneling between two electronic bands
in each superconductor (S) layer.  On the other hand, the conventional 
Josephson effect describes tunneling between two adjacent S layers.  These 
effects determine the dynamics of the phase difference between the condensates
within the same S layer and across two adjacent S layers, respectively.    

The relative phase of the two condensates is fixed in the ground state.  This
relative phase is locked to the value of 0 and $\pi$ when the order parameter
symmetry is $S_{++}$ and $S_{+-}$, respectively.  However, when the  
fluctuations about these phase-locked states are small, the interplay between 
the two Josephson effects can yield interesting phenomena.  One such example is
a collective excitation\cite{Legg} known as the Josephson-Leggett (JL) mode.
This JL mode had been observed\cite{Blum} in $MgB_2$ by Bloomberg and 
coworkers, using Raman scattering.  Also, theoretical studies of 
superconductor-insulator-superconductor junctions between one- and two-gap 
superconductors suggest that the ground state may violate\cite{Ng} the time 
reversal symmetry and that the phase dynamics of LJJ may depend on the symmetry
of the superconducting order parameter.\cite{Ota2,Agt}  This hetero-Josephson 
junction may be fabricated\cite{fab2B} by using $Nb$ (one-gap) and either 
$MgB_2$ or iron-based superconductors (two-gap).  Ota and coworkers suggested 
that the gap symmetry can affect\cite{Ota2} the Josephson current across the 
grain boundaries in polycrystalline samples as well as the current-voltage 
(I-V) characteristics of the multi-gap intrinsic LJJ stacks.  Recent 
theoretical studies of hetero-Josephson junction suggest that the phase 
dynamics of LJJ are affected by the JL mode.  The effects due to availability 
of two tunneling channels in LJJ may appear in measurable physical quantities,
including the drastic enhancement of the macroscopic quantum tunneling rate and
the presence of an extra step structure,\cite{Ota1} in addition to the 
conventional Shapiro steps, in the I-V characteristics.  

A deviation from the phase-locked state in LJJ may not be limited to a small
amplitude.  The experimental data from the magnetic response of a 
superconducting ring with two pseudo-order parameters indicate that a stable 
soliton-shaped phase difference between the two condensates (i.e., $i$-soliton)
is attainable.\cite{Bluhm}  This observation is consistent with a suggestion 
that the phase fluctuations can grow and produce a stable 2$\pi$-phase 
texture.\cite{Tana}  Excitation of $i$-soliton represents large phase 
fluctuations due to the interband Josephson effect and is taken as a hallmark 
of the multi-gap superconductors.  

Soliton states in two-gap superconductors had been explored by a number of authors.  Kupulevaksky and coworkers examined\cite{KOY} this soliton state in
mesoscopic thin-walled cylinders in external magnetic fields by using the 
Ginzberg-Landau approach.  Tanaka and coworkers suggested\cite{TC} that 
interesting excitations, including a phase domain surrounded by the $i$-soliton
wall, can arise in two dimensions (i.e., D=2) since an $i$-soliton may be 
considered as a D-1 dimensional quantum phase dislocation.  The $i$-soliton 
wall may carry a fractional flux quantum when one end of the soliton wall is
terminated by the fractional vortex\cite{Be} while the other end is attached 
to a sample edge.  Also, a vortex-molecule may be formed\cite{GSM} when two 
fractional vortices, with a unit fluxoid quantum as the total magnetic flux, 
are connected by the $i$-soliton bond.   These fractional vortices have been 
observed in a multi-layered superconductor\cite{Luan} by using both magnetic 
force and scanning Hall probe microscopy.

The $i$-solitons differ from the Josephson vortices (i.e., fluxons) since they 
do not carry magnetic flux and do not interact with either magnetic field or 
supercurrents.  However, an $i$-soliton may be formed and driven\cite{GV} by 
nonequilibrium charge density or by sufficiently strong superconducting 
currents.  Gurevich and Vinokur suggested\cite{GV} that spontanteous 
appearance of a soliton-like phase texture represents the breakdown of the 
phase-locked state.  This breakdown can arise when the applied current density
along the superconductor layers exceeds the critical value.  The phase 
fluctuations may appear as either an additional resonances in the AC Josephson
effect or a static $2\pi$-kink in the phase difference.  If the 2$\pi$-phase 
texture exists in each S layer, then this $i$-soliton may change the phase 
dynamics of the LJJ by inducing a critical current density modulation. 

Earlier studies on the effects of critical current modulation indicate that 
both spatial and temporal dependence of Josephson current amplitude may be 
obtained by using experimental techniques such as Ohmic heating, quasiparticle 
injection, and illumination with an intensity modulated beam of 
light.\cite{Vanneste}  Also, a spatially periodic modulation of the critical 
current may be obtained from a periodic array of microresistors in the 
insulator (I) layer.\cite{MS}  As the microresistors behave as pinning centers
for moving Josephson vortices (i.e., fluxons), the speed of the fluxon becomes
modulated near each microresistor.  The effects of a small periodic critical 
current modulation\cite{Vanneste,Mkr,Mal,San} created by an array of 
microresistors can yield a number of interesting properties, including emission
of electromagnetic (EM) radiation and Josephson steps in the I-V 
characteristics.\cite{Vanneste,Mkr}   This suggests that the spatial and 
temporal periodic modulation of the critical current due to excitation of the
$i$-solitons may exibit the same interesting property. 

Emission of EM radiation from an inhomogenous LJJ had been studied by using a 
number of different theoretical approaches, including the Green function 
perturbation technique,\cite{Mkr} inverse scattering perturbation 
theory\cite{Mal} and numerical simulation.\cite{San}  These studies reveal that
a moving fluxon can radiate EM waves when its speed is larger than the critical
value $\upsilon_{th}$.  Also, the interference between the emitted EM waves can
give rise to well-discernible steps in the I-V characteristics.  This suggests
that, if the critical current modulation is generated by large amplitude 
fluctuations of the phase difference, this may affect the junction property 
through the changes in the phase dynamics.  These changes may be measured from 
the I-V curves.  However, the effects of the interband Josephson current on the
phase dynamics of a LJJ with two-gap superconductors have not yet been explored. 

In this paper, we consider a quasi-one dimensional LJJ (i.e. D=1) with the 
dimensions, compared to the Josephson length $\lambda_J$, of $L_x \gg 
\lambda_J$ and $L_y \ll \lambda_J$ (see Fig. 1).  We assume that there are no
fractional vortices and investigate the effects of large phase functuations 
(i.e., $i$-soliton) on the phase dynamics of the LJJ.  We note that, here, a 
weak external magnetic field, applied parallel to the insulator layer, 
penetrates the junction in the form of fluxons.  Before proceeding further, we
outline the main result.  We find that i) large fluctuations in the relative 
phase of the two condensates in each S layer via the interband Josephson effect
may be described by the sine-Gordon equation.  ii) The soliton-like excitation
(i.e., 2$\pi$-phase texture) can generate both a spatial and temporal modulation
of the critical current between the adjacent S layers.  iii) The critical 
current modulation yields emission of EM waves which radiate along the junction
layer in a form of quasi-linear wave.  The strength of the critical modulation 
is characterized in the I-V curve as a discontinous structure.

The outline of the remainder of the paper is as follows.  In Sec. II, we 
describe the phase dynamics of a LJJ with two-gap superconductors by using a set 
of two sine-Gordon equations.  In Sec. III, we derive the equation of motion 
for the relative phase of the two condensates in each S layer.  In Sec. IV, we
discuss the effects of the interband Josephson current on the phase dynamics by
computing the radiation correction to the bare soliton solution.  In Sec. V, we
compute the trajectories of fluxon in the velocity-position phase plane and 
calculate the current-voltage characteristic curves.  Finally, we summarize the
result and conclude in Sec. VI.

\section{Theoretical model for phase dynamics}

In this section, we derive the equation of motion for the phase differences.  
Here, we describe the phase dynamics of the LJJ by neglecting the dissipation
effect, for simplicity, but this effect is included later in the calculation of
the fluxon trajectories in Sec. V.  Also, we neglect the boundary effect by 
considering a region away from the junction boundary in the x-direction (see
Fig. 1), where the Josephson critical current density $J_c^o$ in the absence 
of $i$-soliton is spatially not uniform.\cite{Ustinov}  We discuss the boundary
effect of vanishing Josephson critical current on the I-V curves in Sec. V.

We start with the following model Hamiltonian: 
${\hat {\cal H}}=\sum_\ell{\hat {\cal H}}_{TB,\ell}+{\hat{\cal H}}_T$.  Here the
Hamiltonian ${\hat {\cal H}}_{TB,\ell}$ accounts for the two-gap 
superconductivity in the $\ell$-th S layer, while the Hamiltonian 
${\hat {\cal H}}_T$ accounts for electron tunneling between the two adjacent S 
layers.  First, we consider these Hamiltonian contributions in the absence of
electromagnetic fields.  We write the two-gap Hamiltonian 
${\hat{\cal H}}_{TB,\ell}$ as
\begin{equation}
{\hat {\cal H}}_{TB,\ell}=\int d{\bf r} \left( \sum_{i=s,d} \varepsilon^i
c_{\sigma ,\ell}^{i\dagger}c_{\sigma ,\ell}^i + {\hat {\cal H}}_\ell^{pair} \right)
\label{two-gap}
\end{equation}
where $\varepsilon^i$ describes the energy of electrons in the $i$-band 
($i=s,d$) about the Fermi energy.   For definiteness, we denote the two 
electronic bands involved in superconductivity as $s$ and $d$ bands.  The 
pairing interaction between electrons in the $\ell$-th S layer is described by
the Hamiltonian ${\hat {\cal H}}_\ell^{pair}$ written as
\begin{eqnarray}
{\hat {\cal H}}_\ell^{pair}=-V_{ss}c_{\uparrow ,\ell}^{s\dagger} 
c_{\downarrow ,\ell}^{s\dagger}
c_{\downarrow , \ell}^{s} c_{\uparrow ,\ell}^{s} - V_{dd}c_{\uparrow ,\ell}^{d\dagger} 
c_{\downarrow ,\ell}^{d\dagger} c_{\downarrow ,\ell}^{d} c_{\uparrow ,\ell}^{d}
\nonumber \\
-V_{sd}(c_{\uparrow ,\ell}^{s\dagger} c_{\downarrow ,\ell}^{s\dagger} 
c_{\downarrow ,\ell}^{d} c_{\uparrow ,\ell}^{d} + h.c.)~,~~~~~~~
\end{eqnarray}
where $c_{\sigma ,\ell}^{i \dagger}$ ($c_{\sigma , \ell}^i$) denotes the 
operator which creates (destroys) an electron with spin $\sigma$ in the 
$i$-band, and $V_{ij}$ denotes the strength of pairing interaction between 
electrons in the $i$ and $j$ bands.  On the other hand, the Hamiltonian 
${\hat {\cal H}}_T$ describes the electron tunneling between the two adjacent S
layers that are separated by the I layer.  This Hamiltonian is given by
\begin{equation}
{\hat {\cal H}}_T = \sum_{i\neq j,\sigma} 
(T_{ij} c_{\sigma ,1}^{i \dagger} c_{\sigma , 2}^{j} + h.c.)~,
\end{equation}
where $T_{ij}$ denotes the tunneling matrix element for an electron from the 
$j$ to $i$ band. 

Following Leggett,\cite{Legg}  we consider the subspace spanned by the BCS type
function and introduce pairing operators
\begin{equation}
{\hat \Psi}^{i}_\ell = c_{\uparrow ,\ell}^{i\dagger} c_{\downarrow ,\ell}^{i\dagger}
\end{equation}
to account for the two (i.e., $s$ and $d$) condensates.  Using the eigenvalues 
of the pairing operator $\hat{\Psi}_\ell^i$, we rewrite the Hamiltonian 
${\hat {\cal H}}_{TB,\ell}$ as
\begin{eqnarray}
{\hat {\cal H}_{TB,\ell}} = f_\ell^s (\vert \Psi_\ell^s \vert^2) + 
f_\ell^d (\vert \Psi_\ell^d \vert^2) - V_{ss} \vert \Psi_\ell^s \vert^2 - 
V_{dd} \vert \Psi_\ell^d \vert^2
\nonumber \\
-V_{sd}(\Psi_\ell^{s*}  \Psi_\ell^d + \Psi_\ell^{d*} \Psi_\ell^s ) ~,~~~~~~~~
\label{TBHam}
\end{eqnarray}
where $f_\ell^i(\vert \Psi_\ell^i \vert^2)$ corresponds to the kinetic energy
of the electrons in the $i$-band.  In the absence of contribution from the 
interband pairing (i.e., $V_{sd}=0$), the electrons in the two bands (i.e., $s$
and $d$) do not interact.  The independent one-band gap equation for the
parameter $\Delta^i=V_{ii}\Psi^i$ may be obtained by minimizing Eq. 
(\ref{TBHam}).  We may account for the phase effects of the two condensates by
writing the complex pesudo-order parameter $\Psi_\ell^a$ as
\begin{equation}
\Psi_\ell^{i} = \Psi_{o,\ell}^{i} e^{i\theta_\ell^i}~.
\label{order}
\end{equation}
In this representation, the interband pairing interaction term in Eq. 
(\ref{TBHam}) depends explicitly on the relative phase of the two condensates
(i.e., $s$ and $d$) 
\begin{equation}
\chi_\ell=\theta_\ell^{s} - \theta_\ell^{d}~. 
\end{equation}
The interband pairing contribution describes the Josephson effect within each S
layer due to tunneling of the condensates between the two bands.  Similarly, 
the conditions for the energy extremum yield the two coupled gap equations of
the form
\begin{equation}
\Delta_\ell^i = \sum_{j} V_{ij} \Psi_\ell^j~.
\label{twogap}
\end{equation}
The coupled gap equations of Eq. (\ref{twogap}) have two nontrivial solutions: 
$\chi_\ell =0$ (i.e., $\theta_\ell^s=\theta_\ell^d$) and $\chi_\ell=\pi$ (i.e., 
$\theta_\ell^s=\theta_\ell^d+\pi$) corresponding to the $S_{++}$ and $S_{+-}$
symmetry, respectively. When $V_{sd} > 0$, $\chi_\ell=0$ is the 
stable solution.  On the other hand, when $V_{sd} < 0$, $\chi_\ell = \pi$ is 
the stable solution.

In addition to the phase-locked state, soft modes associated with fluctuations 
of $\chi_\ell$, representing a phase texture, may appear as a 2$\pi$-kink in 
$\chi_\ell$.  These soft modes can modify the phase dynamics of LJJ.  For 
example, these modes can manifest as resonances in the AC Josephson effect when
the two electronic bands are out of equilibrium.  Also, they may appear as 
2$\pi$-kink representing an $i$-soliton. 

\begin{figure}[t]
\includegraphics[width=6.3cm]{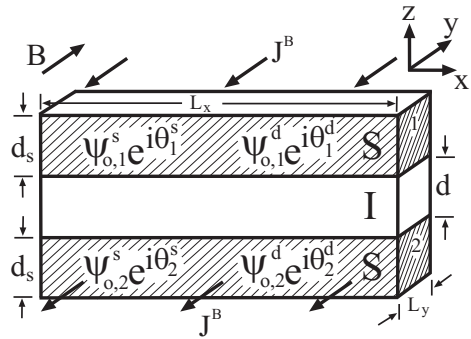}
\caption{A LJJ with two-gap superconductor represented by two pseudo-order 
parameters $\Psi_o^s \exp(i\theta^s)$ and $\Psi_o^d \exp(i\theta^d)$ is shown 
schematically.  Here, $L_x$ and $L_y$ denote the dimensions in the $x-$ and 
$y-$direction, respectively. $J^B$ is the bias current density and $B$ is the 
externally applied magnetic field.  $d_s$ and $d$ denote the thickness of the
superconductor (S) and insulator (I) layer, respectively. }
\label{fig1}
\end{figure}

We now investigate the effects of these soft modes by starting with the
partition function ${\cal Z}={\rm Tr}\exp[-\beta {\cal H}]$ for the LJJ with 
two-gap superconductors\cite{Sh} in the presence of electromagnetic fields
\begin{equation}
{\cal Z} =
\int {\cal D}[A^o, A] {\cal D}[\Psi_o^i ]{\cal D}[\theta^i] 
e^{-{\cal S}[\Psi_o^i , \theta^i, A^o, A]}~,
\end{equation}
where $\beta=1/T$ and $T$ denotes temperature.  Here, we set $\hbar =c= k_B =1$
for convenience.  The scalar and vector potential in the $\ell$-th layer are 
denoted by $A_\ell^o$ and $A_\ell=(A^x, A^y, A^z)_\ell$, respectively.   The 
effective action ${\cal S} = {\cal S}_{gap} + {\cal S}_{field} - {\rm Tr}~
{\hat G}^{-1}$ is obtained by carrying out the imaginary-time 
functional integral over the Grassmann fields  $c_{\sigma ,\ell}^{i \dagger}$ 
and $c_{\sigma ,\ell}^i$.  The contributions from the psudo-order parameter and
electromagnetic fields are denoted by ${\cal S}_{gap}$ and ${\cal S}_{field}$,
respectively.  The expression for ${\cal S}_{gap}$ may be found in appendix and
${\cal S}_{field}$ is included in the phase contribution, ${\cal S}_{phase}$, 
to the action below.   Also, we do not write the explicit expression of the 
inverse Green function ${\hat G}^{-1}$ here but note that it is a 8x8 matrix 
which consists of a 4x4 matrix for each two-gap superconductor layer.  We 
extract the superconducting phase degree of freedom  by performing the unitary transformation for the Green function and follow the standard 
procedure\cite{ASE} of retaining only the second-order tunneling processes arising from the Josephson effect.   The usual imaginary-time functional 
integral approach\cite{Machida} in this approximation allows us to obtain the 
phase contribution ${\cal S}_{phase}$ to the effective action $\cal S$, which
includes the phase terms, in an external magnetic field B applied in the 
y-direction (see Fig. 1) as
\begin{equation}
{\cal S}_{phase} = \int_0^{\beta} d\tau\int dx~ {\cal L}_{p}~.
\label{actphase}
\end{equation}
The effective Lagrangian density ${\cal L}_p$ for the superconducting phases\cite{Machida,LO} is given by
\begin{eqnarray}
{\cal L}_{p} ={d_s \over 8\pi {\bar e}^2} \sum_{i,\ell} \bigg[{1 \over \mu_i^2}
\left({\partial\theta_\ell^{i} \over \partial \tau} + 
{\bar e} A_\ell^o\right)^2 + {1 \over \lambda_i^2}
\left({\partial\theta_\ell^{i} \over \partial x} - 
{\bar e} A_\ell^x\right)^2\bigg] 
\nonumber \\
+\sum_{i,j} {J^{ij} \over {\bar e}} \cos \varphi^{ij}
+ \sum_\ell {J_{inter} \over {\bar e}} \cos \chi_\ell +{\cal L}_{EM}~.~~~~~~~
\end{eqnarray}
where ${\bar e}=2e$, and $d_s$ is the superconductor layer thickness. The phase
difference of the superconducting order parameter in the magnetic field is 
$\varphi^{ij}=\theta_1^i -\theta_2^j - {\bar e} A_{1,2}^z$, where 
$A_{1,2}^z = \int_{\ell=1}^{\ell=2} A^z(z) dz$.  The charge screening length 
$\mu_i = \sqrt{\lambda_{TF}^2/4\pi\epsilon_i}$ is a constant related to the Thomas-Fermi screening length $\lambda_{FT}=\sqrt{\pi a_o /4k_{F}}$, and the 
dielectric constant $\epsilon_i$ of the S layer.  Here, $a_o$ is the Bohr 
radius and $k_F$ is the Fermi vector.  The magnetic penetration depth of the S 
layer are denoted by $\lambda_i = \sqrt{m_i^o/4\pi n_i {\bar e}^2}$ where 
$m_i^o$ is the mass of the $i$-band electron,
\begin{equation}
n_i = \sum_{\bf k}  
{4\varepsilon_{\bf k}^i \over 3E_{\bf k}^i} \left\{
\tanh {{\bar E}_{\bf k}^i \over 2}  + {\bar E}_{\bf k}^i
f( E_{\bf k}^i) [1 - f( E_{\bf k}^i)] \right\} ,
\end{equation}
$E_{\bf k}^i = \sqrt{\varepsilon_{\bf k}^2 +\Delta_{\bf k}^i}$, 
${\bar E}_{\bf k}^i=\beta E_{\bf k}^i$, and $f(E)$ is the Fermi function.  
Here, we make a local approximation and consider $\mu$ and $\lambda$ as 
constants.\cite {ASE, Machida}  The Josephson critical current $J^{ij}$ betwen
the electronic bands $i$ and $j$ in two adjacent S layers may be 
written\cite{ASE, Machida} as
\begin{eqnarray}
J^{ij}=-{2{\bar e} \over d_s\beta}\int d\tau \sum_n e^{-i\omega_n \tau}
\sum_{{\bf k},{\bf k}'} T_{ij}^2 {\Delta_{\bf k}^i \Delta_{\bf k'}^j \over
E_{\bf k}^i E_{\bf k'}^j} \times
\nonumber \\
\Bigg\{ {E_{\bf k}^i - E_{\bf k'}^j \over (E_{\bf k}^i +E_{\bf k'}^j)^2 + 
\omega_n^2} [f(E_{\bf k}^i) - f(E_{\bf k'}^j)] ~~~~~
\\
+ {E_{\bf k}^i + E_{\bf k'}^j \over (E_{\bf k}^i +E_{\bf k'}^j)^2 + 
\omega_n^2} [1- f(E_{\bf k}^i) - f(E_{\bf k'}^j)]
\Bigg\}~, \nonumber
\end{eqnarray}
where $\omega_n = 2n\pi/\beta$ with $n=0, \pm 1, \pm 2, \cdot\cdot\cdot$ is the
Matsubara frequency.  The interband Josephson critical current 
$J_{inter} =2{\bar e} V_{sd} \Psi_o^s\Psi_o^d /(V_{ss}V_{dd}-V_{sd}^2)$ between
the two bands within the same S layer does not represent charge transfer, as in
$J^{ij}$.  The Lagrangian density for the electromagnetic field ${\cal L}_{EM}$
in the insulator layer is given by
\begin{equation} 
{\cal L}_{EM}={d \over 8\pi}[\epsilon (E_{1,2}^z)^2 - (B_{1,2}^y)^2]~,
\end{equation}
where $d$ is the I layer thickness, and $\epsilon$ is the dielectric constant. 
The electric and magnetic field between two adjacent S layers (i.e., $\ell=1$ 
and $\ell=2$) are defined as
\begin{eqnarray}
E_{1,2}^z &=& -{\partial A_{1,2}^z \over \partial t}- 
{1 \over d}(A_1^o - A_2^o) \\
B_{1,2}^y &=& {1 \over d}(A_1^x - A_2^x) - {\partial A_{1,2}^z \over \partial x}
\end{eqnarray}
respectively. 

We now derive the equations of motion for the phase difference $\varphi^{ij}$ 
by following the usual approach of minimizing the action ${\cal S}_{phase}$ of
Eq. (\ref{actphase}).  Here, noting that $\tau =-it$, we examine the phase 
dynamics of both $\varphi^{ss}$ and $\varphi^{dd}$ since we may write that 
$\varphi^{sd} = \theta_1^s -\theta_2^d +{\bar e} d A_{1,2}^z =\varphi^{ss} + 
\chi_2$ and $\varphi^{ds} = \theta_1^d - \theta_2^s + {\bar e}d A_{1,2}^z = 
\varphi^{ss}-\chi_1$.  Using the Euler-Lagrange equation for the scalar and 
vector potentials (i.e.,  $A_\ell^o$ and $A_\ell^z$), we obtain the following
two coupled equations of motion for the phase differences $\varphi^{ss}$ and 
$\varphi^{dd}$: 
\begin{eqnarray}
\eta_\mu^s {\partial^2\varphi^{ss} \over \partial t^2} -  
\eta_\lambda^s {\partial^2\varphi^{ss} \over \partial x^2} -
\zeta_\mu {\partial^2\varphi^{dd} \over \partial t^2} + 
\zeta_\lambda {\partial^2\varphi^{dd} \over \partial x^2}~~~~~~~
\nonumber \\
+{2J^{ss} \over {\bar e}}\sin\varphi^{ss} + {J^{sd} \over {\bar e}} 
\left[ \sin(\varphi^{ss}+\chi_2) +
\sin(\varphi^{ss}-\chi_1) \right]~~~ \\
+ {J_{inter} \over {\bar e}}(\sin \chi_1 - \sin\chi_2 ) = 0~,~~~~~~~~~~~~~
\nonumber
\end{eqnarray}
and
\begin{eqnarray} 
\eta_\mu^d{\partial^2\varphi^{dd} \over \partial t^2} - 
\eta_\lambda^d {\partial^2\varphi^{dd} \over \partial x^2} -
\zeta_\mu {\partial^2\varphi^{ss} \over \partial t^2} + 
\zeta_\lambda {\partial^2\varphi^{ss} \over \partial x^2}~~~~~~~
\nonumber \\
+{2J^{ss} \over {\bar e}^2}\sin\varphi^{dd} + {J^{sd} \over {\bar e}} 
\left[ \sin(\varphi^{dd}-\chi_2) + 
\sin(\varphi^{dd}+\chi_1) \right]~~~ \\
- {J_{inter} \over {\bar e}}(\sin \chi_1 - \sin\chi_2 ) = 0~.~~~~~~~~~~~~~
\nonumber
\end{eqnarray}
The coeffiecients in the above two equations of motion (for $\varphi^{ss}$ and 
$\varphi^{dd}$) are
\begin{equation}
\eta_\xi^i = {d_s \over 4\pi {\bar e}^2\xi_i^2} 
\left( 1-{d_sd \over 4\pi\xi_i^2 \kappa_\xi} \right)
\end{equation}
and
\begin{equation}
\zeta_\xi = \left({d_s \over 4\pi {\bar e} \xi_s\xi_d}\right)^2
{d \over \kappa_\xi},
\end{equation}
where
\begin{equation}
\kappa_\xi = {d_s d \over 4\pi}\left( {1 \over \xi_s^2} +
{1 \over \xi_d^2} \right) + \kappa_\xi^o~.
\end{equation}
Here, $\xi_i$ denotes either the screening length ($\mu_i$) or the magnetic 
penetration depth ($\lambda_i$) of the $i$-band, $\kappa_\mu^o = \epsilon/2\pi$
and $\kappa_\lambda^o = 1/2\pi$.  

By noting that the phase difference $\varphi^{sd}$ may be written as either 
$\varphi^{sd} = \varphi^{ss}+\chi_2$ or $\varphi^{sd}=\varphi^{dd}+\chi_1$, we
may see straightforwardly that the phase dynamics for both $\varphi^{dd}$ and 
$\varphi^{ss}$ are closely related through the soft modes $\chi_1$ and $\chi_2$
in the two adjacent S layers.  We simplify the equation of motion for 
$\varphi^{ss}$ by writing that $\varphi^{dd}=\varphi^{ss}-\chi_1 + \chi_2$.  
This substitution yields a familiar form of the equation of motion for both 
$\varphi^{ss}$ and $\varphi^{dd}$.  Here, we focus on the phase dynamics of 
$\varphi^{ss}$ given by
\begin{eqnarray}
{d_s \over 4\pi {\bar e}^2} \left( 
{\kappa_\mu^o \over \mu_s^2 \kappa_\mu} {\partial^2\varphi^{ss} 
\over \partial t^2}- 
{\kappa_\lambda^o \over \lambda_s^2 \kappa_\lambda }{\partial^2 \varphi^{ss} 
\over \partial x^2} \right) + {2J^{ss} \over {\bar e}} \sin\varphi^{ss} ~~~
\nonumber \\
+ {J^{sd} \over {\bar e}} \left[ \sin(\varphi^{ss}+\chi_2)+
\sin(\varphi^{ss}-\chi_1) \right]~~~~~~~~~~~~~
\label{TBsG} \\
+\sum_{\ell=1}^2 (-1)^{\ell+1} \left[ \zeta_\mu {\partial^2 \chi_\ell \over \partial t^2} 
-\zeta_\lambda {\partial^2\chi_\ell \over \partial x^2} +
{J_{inter} \over {\bar e}} \sin \chi_\ell \right] = 0~.~~~
\nonumber 
\end{eqnarray}
As indicated by Eq. (\ref{TBsG}), the phase dynamics of $\varphi^{ss}$ depend 
strongly on the relative phases (i.e., $\chi_1$ and $\chi_2$) of the two 
condensates.  As the phase $\chi_\ell$ depends only on the interband Josephson 
effect, it is not influenced by the dynamics of the phase difference 
$\varphi^{ss}$ across the junction.  Hence we may separate the phase dynamics 
decribed by Eq. (\ref{TBsG}) into two separate equations.  Assuming that 
$\chi_1 =\chi_2 =\chi$, we write these equations as 
\begin{eqnarray}
{\partial^2 \varphi^{ss} \over \partial {\bar x}^2} -
{\partial^2 \varphi^{ss} \over \partial {\bar t}^2} -
\left( 1 + {J^{sd} \over J^{ss}}\cos \chi \right) \sin\varphi^{ss} = 0~,~
\label{sssGeq}\\
{\partial^2 \chi \over \partial {\tilde x}^2} -
{\partial^2 \chi \over \partial {\tilde t}^2} - \sin\chi = 0~,~~~~~~~~~~~~
\label{isolsG}
\end{eqnarray}
where ($\bar t$, $\bar x$) and ($\tilde t$, $\tilde x$) are the dimensionless 
coordinates for the phases $\varphi^{ss}$, and $\chi$, respectively.  Here, it
is noted that, we may assume that $\varphi^{ss} = \varphi^{dd}$ when 
$\chi_1 =\chi_2$.  This suggests that the two tunneling channels (i.e., 
$\varphi^{ss}=\varphi^{dd}$) become equivalent.   

\section{Interband Josephson effect}

In this section, we derive the equation of motion for the interband phase 
difference $\chi$ of Eq. (\ref{isolsG}) from the two-band Hamiltonian 
${\hat {\cal H}}_{TB,\ell}$ of Eq. (\ref{two-gap}).  It is noted that this Hamiltonian accounts for excitation of an $i$-soliton due to the interband 
Josephson effect.  By following Leggett,\cite{Legg} we write the two-band 
Hamiltonian in the {\bf k}-space representation as
\begin{equation}
{\hat {\cal H}}_{TB,\ell}=\sum_{i=s,d}{\hat {\cal H}}_{OB,\ell}^i - {\hat E}_o + 
{\hat {\cal H}}_{inter,\ell}^{pair} + E_{ng} + {\hat {\cal H}}_\ell^{ci}~,
\label{tbk}
\end{equation}
where the Hamiltonian ${\hat {\cal H}}_{OB,\ell}^i$ for the electronic band $i$
is given by  
\begin{equation}
{\hat {\cal H}}_{OB,\ell}^i = \sum_{{\bf k},\sigma,\ell} \varepsilon_{\bf k}^i
c_{{\bf k}\sigma,\ell}^{i\dagger}c_{{\bf k}\sigma,\ell}^i -
V_{ii} \sum_{{\bf k},{\bf k}'} c_{{\bf k} \uparrow ,\ell}^{i\dagger} 
c_{-{\bf k}\downarrow ,\ell}^{i\dagger}
c_{-{\bf k}'\downarrow , \ell}^i c_{{\bf k}'\uparrow ,\ell}^i .
\end{equation}
Here, we introduce ${\hat E}_o$ so that the ground-state energy of $\sum_{i} 
{\hat {\cal H}}_{OB,\ell}^i - {\hat E}_o$ in the normal state vanishes for an
arbitrary value of $N_s$ and $N_d$, which denote the number of electrons in 
the $s$ and $d$-band, respectively.  The interband pairing interaction between 
the electrons in $s$ and $d$ band is decribed by the Hamiltonian 
${\hat {\cal H}}_{inter,\ell}^{pair}$ which may be written as
\begin{equation}
{\hat {\cal H}}_{inter,\ell}^{pair} = -V_{sd} \sum_{{\bf k},{\bf k}',i\ne j}
c_{{\bf k} \uparrow ,\ell}^{i\dagger} c_{-{\bf k}\downarrow ,\ell}^{i\dagger}
c_{-{\bf k}'\downarrow , \ell}^j c_{{\bf k}'\uparrow ,\ell}^j ~.
\end{equation}
The term $E_{ng}$ denotes the ground-state energy of the system in the normal
state.  This energy $E_{ng}$ is fixed by the total number of electrons 
$N=N_s+N_d$ in the system. The Hamiltonian ${\hat {\cal H}}_\ell^{ci}$ accounts
for the effects of charge fluctuations from the equilibrium state.  These 
charge fluctuations may arise from either the boundary conditions, various 
residual scattering processes, or the chemical potential fluctuations and lead 
to the charge imbalance between the $s$ and $d$-band.  The Hamiltonian 
${\hat {\cal H}}_\ell^{ci}$ may be approximated\cite{Legg} as
\begin{equation}
{\hat {\cal H}}_{\ell}^{ci} = \gamma_o [({\hat N}_s - {\hat N}_s^o) - 
({\hat N}_d - {\hat N}_d^o)]_\ell^2 = \gamma (\delta {\hat N}_\ell)^2
\end{equation}
when the deviation from the equilibrium is small.  Here, 
$\gamma_o =(\rho_s^{-1} + \rho_d^{-1})/8$, $\rho_i=m_i^o k_F/\pi^2$ denotes the
density of states for the $i$-band electrons at the Fermi surface, and 
${\hat N}_i^o$ denotes the number operator for the $i$-band electrons in the 
equilibrium state.

To assess the dynamics of relative phase $\chi_\ell$ due to the interband 
Josephson effect, we rewrite the two-band Hamiltonain ${\hat{\cal H}}_{TB,\ell}$
in terms of the familiar Ginzberg-Landau (GL) free energy.  (See the appendix 
for a detailed discussion on the derivation of GL free energy for the two-gap 
superconuctors.)  Here, we note that the temperature range is restricted to 
$(T_c - T)/T_c \ll 1$, where $T_c$ is the superconducting transition 
temperature.  Also, we considered the zero field limit here since the interband 
phase fluctuations do not carry magnetic flux. In the absence of magnetic 
field, we may write the one-band Hamiltonian ${\hat {\cal H}}^i_{OB,\ell}$ in 
the form of the GL free energy\cite{NG} ${\cal G}_{OB,\ell}^i$ as a function of
pseudo-order parameter $\Phi_\ell^i$ (i.e., 
$\sum_i{\hat {\cal H}}^i_{OB,\ell}-E_o\rightarrow \sum_i{\cal G}_{OB,\ell}^i$).
Here, $\Phi_\ell^i$ is the auxilary field representing an electron pair.  For 
one spatial dimensional case, we may write ${\cal G}_{OB,\ell}^i$ as
\begin{equation}
{\cal G}^i_{OB,\ell}= a^{GL}_i \vert \Phi_\ell^i \vert^2 + 
{b^{GL}_i \over 2}\vert \Phi_\ell^i \vert^4 + {1 \over 2m_i^*} \bigg\vert
{d\Phi_\ell^i \over dx}\bigg\vert^2 ~,
\label{lll1}
\end{equation}
where $a^{GL}_i$ and $b^{GL}_i$ are the coefficients of the GL free energy and 
$m_i^*$ is the effective mass of the electron in the $i$-band. (See appendix.) 
When the order parameter $\Phi_\ell^i$ is written in terms of the modulus-phase
variables [i.e., $\Phi_\ell^i = \vert \Phi_\ell^i \vert \exp(i\theta_\ell^i)$],
the interband pairing interaction ${\hat{\cal H}}_{inter,\ell}^{pair}$ yields 
the contribution
\begin{equation}
{\cal G}_{inter,\ell}^{pair} = -
{2g_{sd} \over g_s g_d} \vert \Phi_\ell^s\vert \vert \Phi_\ell^d\vert 
\cos (\theta_\ell^s - \theta_\ell^d) ~,
\label{lll2}
\end{equation}
where $g_i=(V_{ss}V_{dd} - V^2_{sd})/V_{ii}$ and 
$g_{sd}=V_{sd}(V_{ss}V_{dd} - V^2_{sd})/(V_{ss}V_{dd})$.  Here, we assumed that
the intraband interaction is much larger than the interband interaction (i.e., 
$V_{ij} \ll V_{ii}$).  With this assumption, we may approximate that 
$\vert \Phi_\ell^a \vert \sim \sqrt{N_a}$, which specifies that the intraband 
interaction is independent of the coordinates (i.e.,
$d \vert \Phi_\ell^a \vert /dx \approx 0$).  The charge-imbalance contribution 
${\hat {\cal H}}_\ell^{ci}$ in Eq. (\ref{tbk}) may be expressed in a familiar 
form by using the number-phase uncertainty relationship of
\begin{equation}
[\delta {\hat N}_\ell, {\hat\theta}_{\ell'}^s - {\hat\theta}_{\ell'}^d] =-4i \delta_{\ell\ell'} ,
\end{equation}
where $\delta_{\ell\ell'}$ is the kronecker delta.  Noting that the Heisenberg
equation of motion for the phase difference 
${\hat\chi}_\ell= {\hat\theta}_\ell^s - {\hat\theta}_\ell^d$ is given by
\begin{equation}
{d {\hat\chi}_\ell \over dt} = i[ {\hat\chi}_\ell , {\hat{\cal H}}_{TB,\ell}] ,
\end{equation}
we may write the charge imbalance contribution as
\begin{equation}
{\cal G}_\ell^{ci} = {1 \over 64 \gamma_o} \left( 
{d\chi_\ell \over dt}\right)^2~.
\label{lll3}
\end{equation} 
Combining the result of Eqs. (\ref{lll1}), (\ref{lll2}) and (\ref{lll3}), we 
may write the Gibbs free energy density ${\bar {\cal G}}_{TB} =\sum_i 
({\cal G}_{OB}^i - {\cal G}_{ng}^i) + {\cal G}_{inter}^{pair} + {\cal G}^{ci}$
for the two-gap superconductor in the superconducting state as
\begin{eqnarray}
{\bar {\cal G}}_{TB,\ell} = \sum_i \bigg( a^{GL}_i \vert \Phi_\ell^i \vert + 
{b^{GL}_i \over 2} \vert \Phi_\ell^i \vert^4 +
{\vert\Phi_\ell^i\vert^2 \over 2m_i^*}
\bigg\vert {d\theta_\ell^i \over dx} \bigg\vert \bigg)~
\nonumber \\
+{1 \over 64\gamma_o} \bigg( { d\chi_\ell \over dt} \bigg)^2
-{2 g_{sd} \over g_s g_d} \vert \Phi_\ell^s\vert \vert \Phi_\ell^d\vert 
\cos \chi_\ell~.~~~~
\label{LGfree}
\end{eqnarray}
Here, we assumed that the effects of external fields are weak inside each S 
layer and, thereby, their contributions in ${\bar {\cal G}}_{TB,\ell}$ are
neglected.

We now use the Gibbs free energy of Eq. (\ref{LGfree}) and obtain the equation 
of motion for $\chi_\ell$.  By noting that there is no supercurrent $J_\ell$ 
flowing anywhere inside the bulk superconductor, we write that
\begin{equation}
J_\ell=\sum_i {\vert\Phi_\ell^i \vert \over m_i^*} 
{d\theta_\ell^i \over dx} = 0.
\end{equation}
From the condition that $J_\ell=0$, it is straightforward to obtain that
\begin{equation}
\theta_\ell^d = -{m_d^* \vert \Phi_\ell^s \vert \over 
m_s^* \vert \Phi_\ell^d \vert}
\theta_\ell^s + \delta_o \pi .
\end{equation}
Here, $\delta_o = 0$ for $J_{inter} <0$ ($S_{+-}$ symmetry) and $\delta_o = 1$
for $J_{inter} > 0$ ($S_{++}$ symmetry).  We minimize the Gibbs free energy 
density with respect to the phase variation  (i.e., 
$d{\bar {\cal G}}_{TB,\ell}/d\chi_\ell = 0$) and obtain the equation of motion
for $\chi_\ell$ as
\begin{equation}
{d^2 \chi_\ell \over d{\bar x}^2} - {d^2 \chi_\ell \over d{\bar t}^2}
- \sin \chi_\ell = 0~,
\label{ibsG}
\end{equation}
where the dimensionless coordinates ($\bar x$, $\bar t$) are 
${\bar x} = [4g_{sd}(m_s^* \vert \Phi^d \vert^2 + m_d^* 
\vert \Phi^s \vert^2 )/ g_s g_d \vert \Phi^s \vert \vert \Phi^d \vert]^{1/2} x$
and ${\bar t}=t/(128 \gamma_o g_{sd} \vert \Phi^s \vert \vert \Phi^d\vert/
g_s g_d )^{1/2}$.  It is straightforward to see that Eq. (\ref{ibsG}) may yield
a number of soliton solutions, but, for simplicity, we consider a single-soliton
solution of
\begin{equation}
\chi_\ell ({\bar x}, {\bar t}) = 4 \tan^{-1} \bigg[ \exp \left(\pm {{\bar x} - 
\upsilon_o{\bar t} \over \sqrt{1-\upsilon_o^2}} \right) \bigg] ,
\label{isoliton}
\end{equation}
which is a kink-solution, known as the $i$-soliton.  Here $\upsilon_o$ is the 
speed of $i$-soliton.  The kink-solution of Eq. (\ref{isoliton}) is identical 
to the functional form of the unperturbed fluxon solution.\cite{MS}  However, 
the property of $i$-soliton differs from that of the fluxon.  Unlike fluxon,
an $i$-soliton does not carry a quantum of magnetic flux and cannot be driven 
by the Lorentz force due to the bias current $J^B$ as shown in Fig. \ref{fig1}.  

The equation of motion for $\chi_\ell$ and its solution of Eq. (\ref{isoliton})
indicates that the perturbation effects in the two-gap superconductor may lead
the relative phase $\chi_\ell$ to fluctuate from the phase-locked state of 
$\chi_\ell$.  These phase fluctuations, in turn, yield collective excitations,
but they modify neither the ground state nor the one-particle excitation 
spectrum.  However, if large phase fluctuations representing a 2$\pi$-phase 
texture can be stablized, then the one-particle excitation spectrum may become
modified.

\section{Effects of $i$-solition on phase dynamics}

\begin{figure}[t]
\includegraphics[width=6.3cm]{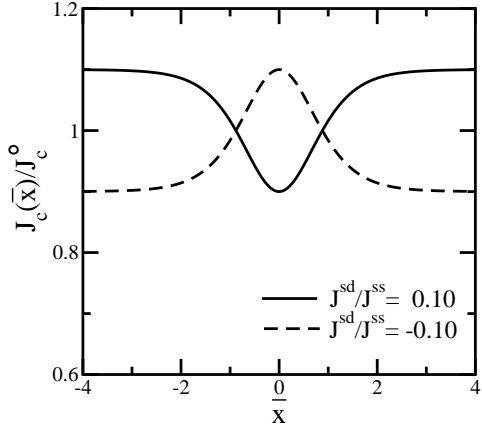}
\caption{The amplitude modulation of the Josephson current density 
$J_c({\bar x})/J_c^o$ due to excitation of static $i$-soliton representing 
2$\pi$-phase texture with its center located at $x=0$ is illustrated for the 
interband Josephson current density of $J^{sd}/J^{ss} =$ 0.10 (solid line) and 
-0.10 (dashed line).  Here, $J_c^o$ is the critical current density of the 
homogenous LJJ in the absence of the interband Josephson effect.}
\label{fig2}
\end{figure}

We now examine the effects of large fluctuations in the relative phase of the
$s$ and $d$ condensates on the fluxon dynamics.  When the amplitude of these
phase fluctuations grows to the nonlinear region and becomes stablized as 
suggested by Tanaka,\cite{Tana} excitation of an $i$-soliton can change the 
amplitude of critical current density.  The modulation of critical current 
between two adjacent S layers may be considered as an effective potential for a
fluxon.  The effects of an $i$-soliton excitation on the phase dynamics of the 
LJJ may be described by Eq. (\ref{sssGeq}).  We may simplify Eq. (\ref{sssGeq})
by substituting Eq. (\ref{isoliton}).   The equation of motion for the fluxon 
which accounts for the interband Josephson effect in each S layer is given as
\begin{equation}
{\partial^2\varphi \over \partial {\bar x}^2} - 
{\partial^2 \varphi \over {\partial {\bar t}^2}}- {J_c \over J_c^o} \sin\varphi = 0 ,
\label{sG1}
\end{equation}
where $\varphi=\varphi^{ss}$, $J_c/J_c^o = 1 + (J^{sd}/ J^{ss}) [1 - 
2\rm{sech}^2 (\alpha_o {\bar x}-\beta_o {\bar t})]$ and $J_c^o$ is the critical
current in the absence of the interband Josephson effect (i.e., $J^{sd}=0$).
Here $\alpha_o=[J^{sd}\lambda_d^2/(1-\upsilon_o^2)J^{ss} d_s d]^{1/2}$ and 
$\beta_o = \upsilon_o [J^{sd}\mu_d^2 \epsilon/(1-\upsilon_o^2) J^{ss} d_s 
d]^{1/2}$.  Equation (\ref{sG1}) indicates that a $i$-soliton, representing a 
moving 2$\pi$-phase texture of Eq. (\ref{isoliton}), leads to both spatial and
temporal dependent modulation of the critical current.  

The effects of spatial and temporal variation of $J_c/J_c^o$ on the fluxon 
depend on the shape of this modulation.  In Fig. \ref{fig2}, we neglect the 
temporal modulation and plot the amplitude of the critical current ($J_c/J_c^o$)
versus the dimensionless spatial coordinate $\bar x$ for $J^{sd}/J_c^o$=0.1 
(solid line) and -0.1 (dashed line) to illustrate the effects of a single
$i$-soliton excitation in the $S_{++}$ and $S_{+-}$ symmetry superconductor, 
respectively.  Here, we set $\alpha_o =1$ for definiteness.  The curves show 
that the shape of critical current modulation depends on the symmetry of order
parameter (i.e., $S_{++}$ versus $S_{+-}$).  However, as we discuss in Sec. V,
when $J^{sd}/J^{ss} \ll 1$, the symmetry of the order parameter does not affect
the fluxon motion significantly.

The critical current modulation induced by the interband Josephson effect has 
two main effects.  First, the shape of fluxon may become deformed.  However,
for a small modulation considered in the present work, this effect is 
negligible.  Second, the speed of fluxon becomes modified since the critical 
current modulation behaves as an effective potential.  In the region of 
critical current modulation, the fluxon speed may become significantly changed 
from a uniform value.  These changes imply that the EM waves can be emitted by 
a moving fluxon as it decelerates.  

We use the perturbation method to examine the effects of critical current 
modulation on the emission of EM waves.  To obtain physical insight, we carry 
out the calculation in the rest frame of the fluxon (i.e., a reference frame
which is moving with the speed of the unperturbed fluxon) as 
described\cite{Fogel} by Fogel and coworkers.  To this end, we perform the 
Lorentz transformation of
\begin{equation}
{\bar t}' = {{\bar t} - \upsilon {\bar x}\over \sqrt{1-\upsilon^2}}
 ~~~~{\rm and} ~~~~
{\bar x}' = {{\bar x} - \upsilon {\bar t} \over \sqrt{1-\upsilon^2}}~,
\end{equation}
where $\upsilon$ is the speed of the unperturbed fluxon.   With this 
transformation, we rewrite the sine-Gordon equation of Eq. 
(\ref{sG1}) as
\begin{equation}
{\partial^2\varphi \over \partial {\bar x}'^2} - {\partial^2 \varphi \over 
{\partial {\bar t}'^2}} - {{\bar J}_c \over J_c^o} \sin\varphi = 0~,
\label{sGbar}
\end{equation}
where ${\bar J}_c/J_c^o = 1 + (J^{sd}/J^{ss}) [1 - 2\rm{sech}^2 (\alpha'_o 
{\bar x}'+ \beta'_o {\bar t}')]$, $\alpha'_o = (\alpha_o - \beta_o \upsilon)/ 
\sqrt{1-\upsilon^2}$ and $\beta'_o =(\alpha_o \upsilon - \beta_o)/ 
\sqrt{1-\upsilon^2}$.  

By considering the case of weak interband Josephson effect (i.e. 
$J^{sd}/J^{ss} \ll 1$), we assume that a solution to Eq. (\ref{sGbar}) may be 
written as 
\begin{equation}
\varphi (x,t)\approx\varphi_o (x) + {J^{sd} \over J^{ss}}\varphi_1(x,t)~.
\end{equation}
Here, for convenience, we make the following changes in the notation: 
$({\bar x}',{\bar t}') \rightarrow (x,t)$.  A solution to the unperturbed 
sine-Gordon equation of 
\begin{equation}
{\partial^2 \varphi_o \over \partial x^2} - 
{\partial^2 \varphi_o \over \partial t^2} -\sin \varphi_o = 0
\end{equation}
is given by $\varphi_o (x) = 4\tan^{-1}[\exp(x)]$.  Following Fogel et 
al.,\cite{Fogel} we may write the correction term $\varphi_1$ due to the 
critical current modulation in the most appropriate basis by separating the 
spatial and temporal dependence as 
\begin{equation}
\varphi_1(x,t) = f(x) e^{-i\omega t}~.
\end{equation}
The separation of variables for the perturbative contribution $\varphi_1$ in the
rest frame of the fluxon (i.e., $\upsilon =0$) leads to the eigenvalue equation
for $f(x)$ as
\begin{equation}
\bigg[ -{d^2 \over dx^2} + (1-2{\rm sech}^2 x) \bigg] f(x) =\omega^2 f(x)~.
\label{sGeigen}
\end{equation}
The eigenvalue problem of Eq. (\ref{sGeigen}) yields one bound state with 
$\omega = \omega_b = 0$ and a continuum of scattering states with 
$\omega^2=\omega_{\kappa}^2 = 1+\kappa^2$.  The corresponding normalized 
eigenfunctions are
\begin{equation}
f(x) = \cases{ f_b(x)=2{\rm sech} x~, & with $\omega=\omega_b$ \cr
               f(\kappa,x) ={\kappa + i {\rm tanh} x \over \sqrt{2\pi}\omega_\kappa} 
               e^{i\kappa x}, & with $\omega=\omega_\kappa$ \cr}
\label{eigenf}
\end{equation}
Here we use the subscripts $b$ and $\kappa$ to denote the bound state and 
continuum of scattering state $\kappa$, respectively.  The bound state $f_b(x)$
is associated with the Goldstone translation mode of the fluxon, while the 
continuum eigenfunctions $f(\kappa,x)$ represent the radiation modes.  The 
eigenfunction of Eq. (\ref{eigenf}) indicates that the first order correction 
$\varphi_1(x,t)$ due to the critical current modulation may be separated into 
two parts as
\begin{equation}
\varphi_1(x,t) = \varphi_{trans}(x,t) + \varphi_{rad}(x,t)~.
\end{equation}
Here, $\varphi_{trans}$ and $\varphi_{rad}$ represent the bound state and 
continuum eigenstate contribution, respectively.  

The bound state contribution $\varphi_{trans}(x,t)$ may be written as  
\begin{equation}
\varphi_{trans}(x,t) = {1 \over 8} \phi_b(t) f_b(x)~.
\label{transcorr}
\end{equation}
The amplitude $\phi_b(t)$ of the bound state is determined straightforwardly 
from the equation of
\begin{equation}
{d^2\phi_b(t) \over dt^2} = 4 \int_{-\infty}^{\infty} dx (1-2{\rm sech}^2
\xi_o ) {{\rm sinh}x \over {\rm sech}^2 x}
\label{bound}
\end{equation}
where $\xi_o = \alpha_o' x + \beta_o' t$.  The solution to Eq. (\ref{bound}) 
may be obtained as
\begin{equation}
\phi_b (t) = -{8 \alpha_o' \over \beta_o'^2}  \left( 1 - \int_{-\infty}^{\infty} dx  
{{\rm sech}^2x \over e^{2\beta_o' t} e^{2\alpha_o' x} +1} \right) .
\end{equation}
We note that $\phi_b(t)$ may be used to evaluate the translation mode 
contribution $\varphi_{trans} (x,t)$.  This contribution has no effects on the 
motion of the fluxon center. 

\begin{figure}[t]
\includegraphics[width=6.3cm]{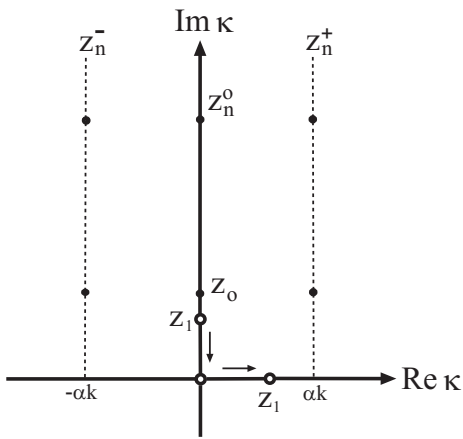}
\caption{The pole structure for the radiation contribution of Eq. 
(\ref{radcorr}) to $\varphi_1(\kappa, t)$ for a fixed $k$ is shown 
schematically.  The solid circles represent the poles yielding the 
exponentially localized contribution.  The open circles represent the changes 
in the location of $z_1$ as the fluxon velocity $\upsilon$ changes 
from $\upsilon < \upsilon_{th}$ (on the imaginary axis) to 
$\upsilon >\upsilon_{th}$ (on the real axis).  The shift direction for the pole
$z_1$ is indicated by the arrows. }
\label{poles}
\end{figure}

The continuum eigenstate contribution, representing the radiation modes, is 
given by
\begin{equation}
\varphi_{rad}(x,t)=\int_{-\infty}^{\infty} d\kappa~\phi(\kappa,t) f(\kappa, x).
\label{radcorr}
\end{equation}
The amplitude $\phi(\kappa , t)$ is determined from
\begin{equation}
{d^2\phi(\kappa,t) \over dt^2} + (1+\kappa^2) \phi(\kappa, t) = 
{\cal Q}(\kappa , t)~,
\label{sGrad}
\end{equation}
where
\begin{equation}
{\cal Q}(\kappa , t) = 2\int_{-\infty}^{\infty} dx f^*(\kappa , x)
(1-2{\rm sech}^2 \xi_o) {{\rm sinh}x \over {\rm cosh}^2 x} .
\label{contik}
\end{equation}
In obtaining Eq. (\ref{sGrad}), we used the orthonormality condition for the
eigenfunctions (i.e., $\int dx f^*(\kappa' ,x) f(\kappa , x)=\delta
(\kappa -\kappa')$).  The contribution to the radiation mode of $\varphi_1$ may
be estimated by solving Eq. (\ref{sGrad}).  For a single modulation of the 
critical current density as shown in Fig. \ref{fig2}, we may obtain a solution 
to Eq. (\ref{sGrad}) more easily by using the relation
\begin{equation}
{\rm sech}^2 \xi_o = \int_{-\infty}^{\infty} {dk \over 2\pi}
{\pi k \over {\rm sinh}{\pi k \over 2}} e^{ik\xi_o} 
\end{equation}
which is the Fourier representation of the critical current variation.  Using 
this substitution, we rewrite ${\cal Q}(\kappa , t)$ by integrating the right 
hand side of Eq. (\ref{contik}) over $x$ and obtain
\begin{eqnarray}
{\cal Q}(\kappa, t) ={-i\pi \over \sqrt{2\pi(1+\kappa^2)}} \bigg \{
(1+ \kappa^2) {\rm sech}{\kappa\pi \over 2} -~~~~~~~
\nonumber \\
\int_{-\infty}^{\infty} dk {{\rm sech}{\pi k \eta_\kappa \over 2} 
\over 2~{\rm sinh}{\pi k \over 2}} 
[1+ \kappa^2 -(k\alpha)^2] k e^{ik \beta'_o t} \bigg\},
\end{eqnarray}
where $\eta_\kappa=\alpha'_o -(\kappa /k)$.  The solution $\phi(\kappa,t)$
may be written as
\begin{equation}
\phi(\kappa , t) =\int_{-\infty}^{\infty} {d\omega \over 2\pi}
{{\cal Q}(\kappa , \omega) \over (1+\kappa^2) - \omega^2} 
e^{i\omega t}  .
\end{equation}
where ${\cal Q}(\kappa ,\omega) = \int dt' {\cal Q}(\kappa , t')
\exp(-i\omega t')$.  It is straightforward to evaluate the integration over $t'$ and $\omega$, and write the solution $\phi(\kappa, t)$ as
\begin{eqnarray}
\phi(\kappa, t) = {-i\pi \over \sqrt{2\pi(1+\kappa^2)}} \bigg[
{\rm sech}{\kappa \pi \over 2} -~~~~~~~~~~~~~~~~~~~
\nonumber \\
~~\int_{-\infty}^{\infty} dk {{\rm sech} {\pi k \eta_\kappa \over 2}
\over {\rm sinh}{\pi k  \over 2}}{{1+ \kappa^2 -(k \alpha_o')^2 \over
1+\kappa^2-(k \beta_o')^2} } k e^{ik \beta_o' t} \bigg]~
\end{eqnarray}
However, as indicated in Eq. (\ref{radcorr}), we need to integrate over the 
continuum variable $\kappa$ to compute $\varphi_{rad}(x,t)$.  This may be 
evaluated by using the contour integration method.  The location of the poles 
for the contour integral is shown schematically in Fig. \ref{poles}.  The 
radiation contribution $\varphi_{rad}(x,t)$ of Eq. (\ref{radcorr}) indicates 
that all poles are simple, and the residue of each pole may be evaluated 
separately.  The location of the poles are the following: $z_o = +i$, 
$z_1 = +i\sqrt{1-(k\beta'_o)^2}$, $z_n^o = +i(2n+1)$, and 
$z_n^\pm = \pm k\alpha'_o + i(2n+1)$, where $n=0, 1, 2, \cdot\cdot\cdot$.  The
residues of the pole structure shown in Fig. \ref{poles} yield two types of 
contribution: i) exponentially localized contribution around the fluxon center 
and ii) linear traveling wave contribution.  Hence, we may decompose the 
radiation mode $\varphi_{rad}$ into the exponentially localized 
($\varphi_{rad}^{exp}$) and traveling wave ($\varphi_{rad}^{wave}$) 
contributions: $\varphi_{rad}=\varphi_{rad}^{exp} + \varphi_{rad}^{wave}$.  We 
note that the exponentially localized contribution $\varphi_{rad}^{exp}$ does 
not produce a true radiative correction.  Only the traveling wave 
$\varphi_{rad}^{wave}$ gives rise to a true radiative contribution.

We now focus on the poles that give rise to the traveling wave contribution to
the radiation correction.  As indicated by the Fourier components of the 
critical current modulation described by the ($1-2{\rm sech}^2\xi_o$) factor in 
Eq. (\ref{contik}), the condition for the traveling wave radiative contribution 
depends on $k$.  For fixed $k < (1/\beta_o')$, the pole at $z_1$ lies on the 
imaginary axis as shown in Fig. \ref{poles}.   However, as the fluxon speed 
$\upsilon$ increases, the pole $z_1$ moves down the imaginary axis.  At the 
critical value $\upsilon=\upsilon_{th}$, the pole $z_1$ lies at the complex 
plane and it becomes real when $\upsilon > \upsilon_{th}$.  The changes in the
radiation contribution in Eq. (\ref{radcorr}) from this pole may be easily 
identified by the contour integration since it is not exponentially localized 
around the fluxon center but oscillates with $x$.  The oscillatory 
contribution only arises when $\upsilon > \upsilon_{th}$.  This leads to the 
radiative contribution of
\begin{eqnarray}
\varphi_{rad}^{wave} (x,t) = -\int_{1 \over \beta_o'}^{\infty} {dk \over 2 \eta} 
{\pi k \over {\rm sinh}{\pi k \over 2}} (\eta + i {\rm tanh} x) \times~~~~
\nonumber \\
\left( 1 - {\alpha_o'^2 \over \beta_o'^2} \right) \left[
{e^{ik(\beta_o' t + \eta x)} \over {\rm cosh}{\pi k \eta_- \over 2}} +
{e^{-ik(\beta_o' t - \eta x)} \over {\rm cosh}{\pi k \eta_+ \over 2}}\right]~,~
\label{radosc}
\end{eqnarray}
where $\eta=\sqrt{(k\beta_o')^2-1}$, and $\eta_\pm = \alpha_o' \mp (\eta/k)$.  Equation 
(\ref{radosc}) indicates that, for a fixed $k$, this radiation correction is 
the superposition of two linear traveling waves with different amplitudes.  The
two waves travel in opposite directions.  The threshold velocity 
$\upsilon_{th}$ for the fluxon is given by
\begin{equation}
\upsilon_{th} = \upsilon_o \left( {\mu_d^2 \epsilon \over \lambda_d^2} \right)^{1/2}~.
\end{equation}
The dependence of $\upsilon_{th}$ on the $i$-soliton velocity $\upsilon_o$ 
indicates that, for the case of static spatial variation of the phase
(i.e., $\upsilon_o=0$), EM radiation may be emitted by the fluxon whenever 
it passes through a region where the critical current is affected by the 
interband Josephson effect.  Hence, when an array of static $i$-solitons are 
excited to yield a spatially periodic modulation of the critical current 
density, the threshold velocity $\upsilon_{th}$ becomes finite.  This radiative
threshold is similar to that found in earlier studies.\cite{Mkr, Mal, San}

\section{Current-voltage characteristics}

Emission of EM radiation by a moving fluxon, indicated by the linear traveling
wave contribution to the radiative correction in $\varphi_1(x,t)$ as discussed 
in Sec. IV, reflects the changes in the fluxon dynamics.  We now examine the 
effects of a single $i$-soliton excitation on fluxon dynamics by considering a
static phase texture described by Eq. (\ref{sG1}) with $\upsilon_o=0$ (i.e., 
$\beta_o =0$).  Here, we include the perturbative effects of bias current and 
dissipation to examine a realistic tunnel junction.  The bias current $J^B$ 
acts as a driving force for the fluxon, while the two dissipation terms, 
$\Gamma_1(\partial \varphi/\partial {\bar t})$ and 
$\Gamma_2 (\partial^3 \varphi/ \partial {\bar t} \partial^2 {\bar x})$, account
for the interaction between the fluxon and dissipative environment.  Using the
perturbed sine-Gordon equation with the critical current density modulation, we
determine the fluxon trajectories and estimate the effects of the bias current 
and dissipation ($\Gamma_1$ and $\Gamma_2$).  These perturbation contributions
as well as the interband Josephson effect can modify the current-voltage curve.
We start with the perturbed sine-Gordon equation of
\begin{equation}
{\partial^2\varphi \over \partial {\bar x}^2} - {\partial^2 \varphi \over 
{\partial {\bar t}^2}} - \sin\varphi = {\cal F}(\varphi , {\bar x}, {\bar t})
\label{sGeqreal}
\end{equation}
where the perturbation term ${\cal F}(\varphi , {\bar x}, {\bar t})$ is given by
\begin{eqnarray}
{\cal F}(\varphi , {\bar x}, {\bar t}) = 
{J^{sd} \over J^{ss}} [1 - {\rm{sech}}^2 (\alpha_o {\bar x})] \sin\varphi - 
{J^B \over J_c^o}~~~~~
\nonumber \\
+\Gamma_1{\partial\varphi \over \partial {\bar t}} 
+\Gamma_2{\partial^3\varphi \over \partial {\bar t}\partial^2 {\bar x}}~.~~~~~~~~~~
\label{pertur}
\end{eqnarray}
Here, we assume that each perturbation term in $\cal F$ is small and does not 
change the shape of the fluxon in the leading order.  The main effect of the 
first term of $\cal F$ in Eq. (\ref{pertur}) is to provide a potential for a
moving unperturbed fluxon of
\begin{equation}
\varphi_o ({\bar x}, {\bar t})= 4\tan^{-1}(e^{\zeta})~,
\label{soliton}
\end{equation}
where
\begin{equation}
\zeta ({\bar x},{\bar t}) = \pm {{\bar x} - \int_0^{\bar t} \upsilon(t') dt' 
-{\bar x}_o({\bar t}) \over \sqrt{1 - \upsilon^2({\bar t})}}~.
\label{trajarg}
\end{equation}
Here, the fluxon speed $\upsilon ({\bar t})$ accounts for the time dependence
induced by the critical current modulation.  We now examine the trajectories 
of the fluxon in an LJJ, and the effects of an $i$-soliton excitation on the
I-V curve in the low-voltage regime.

By following McLaughlin and Scott,\cite{MS} we examine the fluxon trajectories
by computing the fluxon speed and the corresponding position.  For this 
purpose, we assume that the fluxon approaches the region of a large stable 
variation of phase difference (i.e., 2$\pi$-phase texture) between the $s$ and 
$d$ condensates  from ${\bar x}=-\infty$ with increasing $\bar t$.  We write 
Eq. (\ref{sGeqreal}) as two first order differential equations describing the
velocity ($\upsilon$) and the position (${\bar x}_o$) of the fluxon, 
respectively, as
\begin{eqnarray}
{d\upsilon \over d{\bar t}} &=& \mp {1-\upsilon^2 \over 4} \int_{-\infty}^{\infty}
d{\bar x}~{\cal F}(\varphi_o , {\bar x}, {\bar t})~{\rm sech}\zeta~,
\label{ode1}
\\
{d{\bar x}_o \over d{\bar t}} &=& -{\upsilon \over 4} \sqrt{1-\upsilon^2} 
\int_{-\infty}^{\infty}d{\bar x}~ {\cal F}(\varphi_o , {\bar x}, {\bar t})~ 
\zeta {\rm sech} \zeta ,
\label{ode2}
\end{eqnarray}
where $\zeta = \zeta({\bar x} ,{\bar t})$ is defined in Eq. (\ref{trajarg}).  
The perturbation term $\cal F$ modifies the speed $\upsilon$ of the unperturbed
wave form of Eq. (\ref{soliton}) to depend on time $\bar t$.  By performing the
integration, we rewrite Eqs. (\ref{ode1}) and (\ref{ode2}), which 
respectively account for the fluxon speed $\upsilon$ and position $X$, as
\begin{eqnarray}
{d\upsilon \over d{\bar t}} = \pm{\pi \over 4} {J^B \over J_c^o} 
(1-\upsilon^2)^{3 \over 2} - 
\Gamma_1 \upsilon (1-\upsilon^2) - {1 \over 3}\Gamma_2\upsilon ~~~~~~~
\nonumber \\
~~~~~~~~~~~~~~
+ (1-\upsilon^2)^{3 \over 2} {J^{sd} \over J^{ss}} 
\int_0^\infty dy \Upsilon_+ {{\rm sinh}y \over {\rm cosh}^3 y}~~~~~~~~
\label{traj1}
\end{eqnarray}
and
\begin{equation}
{dX \over d{\bar t}} = \upsilon + {\upsilon -\upsilon^3 \over 2}
{J^{sd} \over J^{ss}}
\bigg( 1-2\int_0^\infty dy \Upsilon_- {y~{\rm sinh}y \over {\rm cosh}^3y} \bigg)~,~~
\label{traj2}
\end{equation}
where $\Upsilon_\pm = \Upsilon_\pm(y,{\bar t},X)= {\rm sech}^2\alpha_o 
(\sqrt{1-\upsilon^2}y+X) \pm {\rm sech}^2 \alpha_o (\sqrt{1-\upsilon^2}y-X)$ 
and $X=\int_0^{\bar t}\upsilon(t') dt' + {\bar x}_o({\bar t})$.  These two 
equations describe the fluxon trajectories in the ($\upsilon$, $X$) phase plane.
We numerically integrate Eqs. (\ref{traj1}) and (\ref{traj2}) to estimate the 
fluxon trajectories.  The uniform fluxon speed $\upsilon = \upsilon_\infty$ 
far away from the region of critical current modulation is given by
\begin{equation}
{d \upsilon \over d{\bar t}} =\pm {\pi \over 4} {J^B \over J_c^o} 
(1-\upsilon^2)^{3 \over 2}
-\Gamma_1 \upsilon(1-\upsilon^2) -{1 \over 3} \Gamma_2 \upsilon .
\end{equation}
The power-balance velocity $\upsilon_\infty$ may be estimated by setting 
$d\upsilon/d{\bar t} =0$.  The fluxon speed $\upsilon_\infty$ is obtained by 
solving the following cubic equation:
\begin{equation}
\sum_{i=0}^3 a_iz^i=0~,
\end{equation}
where $z = (1-\upsilon_\infty)^{1/2}$, $a_3 = [\pi^2 (J^B)^2/16(J_c^o)^2] + 
\Gamma_1^2$, $a_2=(2\Gamma_1\Gamma_2/3)-\Gamma_1^2$, $a_1 = (\Gamma_2^2/9) - 
(2\Gamma_1\Gamma_2/3)$ and $a_0= -\Gamma_2^2/ 9$.   We note that the solution 
is bounded by the condition that $0 \le \upsilon_\infty \le 1$ since 
$\upsilon_\infty$ is given in units of Swihart velocity.

\begin{figure}[t]
\includegraphics[width=6.3cm]{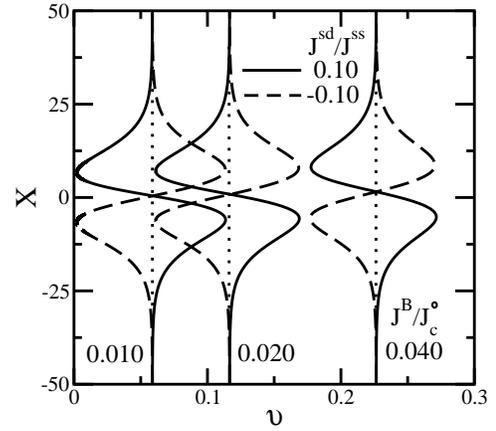}
\caption{Fluxon trajectories in the ($\upsilon$, $X$) phase plane for 
$\Gamma_1 =\Gamma_2 =0.1$ are plotted to illustrate their dependence on the 
strength of both the bias current $J^B/J_c^o$ and interband Josephson curent 
$J^{sd}/J^{ss}$.  Three solid (dashed) curves, from left to right, correspond 
to $J^B/J_c^o =$ 0.010, 0.020 and 0.040 for $J^{sd}/J^{ss} =$ 0.10 (-0.10). 
The vertical dotted lines represent the uniform fluxon speed in the absence of
critical current modulation.} 
\label{fig4}
\end{figure}

We now estimate the effects of bias current $J^B/J_c^o$ on the fluxon 
trajectories in the ($\upsilon$, $X$) phase plane.  In Fig. \ref{fig4}, the 
fluxon trajectories obtained by numerically integrating Eqs. (\ref{traj1}) and
(\ref{traj2}) are plotted for $J^B/J_c^o =$ 0.01, 0.02, and 0.04 (from left to 
right) to illustrate the position $X$ of the fluxon as a function of velocity
$\upsilon$.  Here, we set the dissipation parameters $\Gamma_1 =\Gamma_2 =$0.1.
The solid (dashed) curves represent the interband Josephson current density $J^{sd}/J^{ss}=$ 0.1 (-0.1).  Here, $\upsilon_\infty$ is the uniform initial 
speed of the fluxon at a position far away from the region of 2$\pi$-phase 
texture which is centered at $X = 0$.  The value of $\upsilon_\infty$ depends 
on $J^B/J_c^o$.  The curve for each $J^B/J_c^o$ shows that, as the fluxon 
approaches $X = 0$, the fluxon speed deviates from a straight vertical dotted 
line representing a uniform speed in the absence of the critical current 
modulation.  The curves also show that when the bias current is small (i.e., 
see, for example, $J^B/J_c^o =$ 0.010) the fluxon becomes
almost pinned since the pinning effects of the critical current modulation 
reduce the fluxon speed close to zero. For $J^B/J_c^o < 0.100$, the fluxon is 
pinned: $\upsilon = 0$ at $X \approx 6.5$ for $J^{sd}/J^{ss} = 0.10$ and
$X \approx -6.5$ for $J^{sd}/J^{ss} = -0.10$.  However, for $J^B/J_c^o >$ 
0.010, the fluxon is not pinned since the driving force due to the bias current
is larger than the pinning force.  The fluxon, approaching from $X=-\infty$, 
undergoes a notable change in its speed in the region of the critical current 
modulation.  The difference between the maximum and minimum value of the fluxon
speed becomes reduced with the increasing bias current density.  This is due to
the fact that the importance of the pinning effect decreases with increasing
value of $\upsilon_\infty$ (i.e., increasing $J^B/J_c^o$).  

As its speed decreases, the fluxon radiates EM waves in the region of critical
current modulation.  A variation in the fluxon sped may be seen easily in Fig. 
\ref{fig4} as a deviation of the fluxon trajectories from the vertical dotted
lines.  For the solid lines representing the $S_{++}$ symmetry (i.e., 
$J^{sd}/J^{ss} > 0$), the fluxon speed $\upsilon \geq \upsilon_\infty$ for 
$X < 0$, but $\upsilon \leq \upsilon_\infty$ for $X > 0$.  On the other hand,
for the dashed lines representing the $S_{+-}$ symmetry (i.e., 
$J^{sd}/J^{ss} < 0$), the fluxon speed $\upsilon \leq \upsilon_\infty$ for 
$X < 0$, but $\upsilon \geq \upsilon_\infty$ for $X > 0$.  The deviation from 
the uniform fluxon speed $\upsilon_\infty$ increase with the interband 
Josephson current density $\vert J^{sd}/J^{ss} \vert$.  This non-uniform fluxon
speed due to the critical current modulation suggests that the fluxon absorbs 
energy from the interband Josephson current to increase the speed from 
$\upsilon_\infty$, but it returns the energy back to the junction by radiating 
EM waves, resulting the decrease in the speed.  

\begin{figure}[t]
\includegraphics[width=6.8cm]{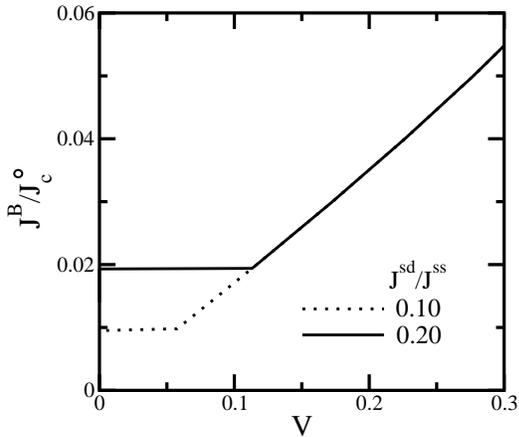}
\caption{The current-voltage curves for $J^{sd}/J^{ss}=$ 0.10 (dotted line) and
0.20 (solid line) illustrate the dependence of the threshold bias current 
(i.e., the value of $J^B/J_c^o$ for $V=0$) on the interband Josephson current.
 Here, the dissipation parameters are $\Gamma_1 = 0.10$ and $\Gamma_2 = 0.10$.} 
\label{fig5}
\end{figure}

As suggested by the fluxon trajectories in the ($\upsilon$, $X$) phase plane,
the bias current density must be larger than the threshold value $J_{th}^B$ in
order for the fluxon to pass through the region of critical current modulation.
In Fig. \ref{fig5}, we plot the I-V curve for $J^{sd}/J^{ss}=$ 0.10 (dotted 
line) and 0.20 (solid line) to illustrate the dependence of the threshold bias
current $J^B_{th}$ on the interband Josephson current ($J^{sd}/J^{ss}$).  The 
fluxon speed, according to the AC Josephson effect 
$(1 /i)(\partial \varphi /\partial t)=V$, is proportional to the voltage $V$ 
across the junction.  Here, we set the dissipation parameters as
$\Gamma_1 = \Gamma_2 =$ 0.1 for concreteness.  The value of the threshold bias
current $J_{th}^B$ for a fixed $J^{sd}/J^{ss}$ may be estimated easily since 
the voltage across the LJJ does not appear until the bias current reaches the
threshold value.  The curves show that the threshold bias current increases 
with  $J^{sd}/J^{ss}$.  As the critical current modulation plays the role of an
effective potential for the fluxon, a larger $J^B/J_c^o$ is needed to overcome 
the pinning effect as $J^{sd}/J^{ss}$ increases.  Hence the threshold bias 
current is similar to the minimum current density needed to overcome the 
pinning force.  In Fig. \ref{fig5}, for  $J^B > J^B_{th}$, the I-V curves show 
that the voltage increases steadily with increasing $J^B$ because the boundary 
effects are not included in this work.  The zero-field step resonances due the
boundary current\cite{Ustinov} (i.e., a rapidly vanishing critical current near
the edges of the junction) are expected to be present along with a smooth 
increase shown in Fig. \ref{fig5} when the boundary effects are included. 

\begin{figure}[t]
\includegraphics[width=6.8cm]{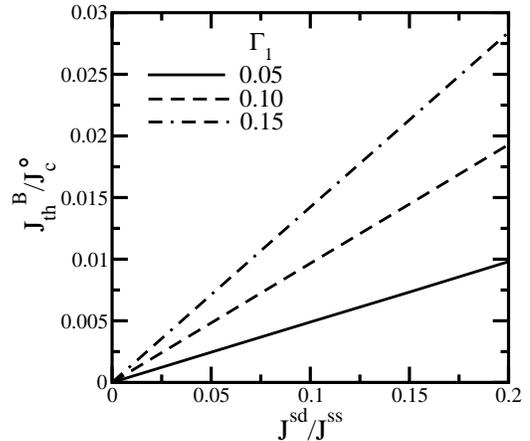}
\caption{The threshold bias current density $J^B_{th}/J_c^o$ is plotted as a 
function of $J^{sd}/J^{ss}$ for $\Gamma_1 =$ 0.05 (solid line), 0.10 
(dashed line), and 0.15 (dot-double dashed line) to illustrate the minimum 
bias current needed to overcome the pinning effect of the critical current 
modulation.  Here, $\Gamma_2 =$ 0.10. }
\label{fig6}
\end{figure}

In Fig. \ref{fig6}, we plot $J^B_{th}/J_c^o$ as a function of $J^{sd}/J^{ss}$
for $\Gamma_1=$ 0.05 (solid line), 0.10 (dashed line) and 0.15 (dot-double 
dashed line) to illustrate the dependence of the threshold bias current on the 
interband Josephson effect.  We set $\Gamma_2=$ 0.10 for concreteness, but the 
curves are independent of the strength of the parameter $\Gamma_2$.  The curves
show that $J_{th}^B$ increases linearly with the interband Josephson current 
$J^{sd}/J^{ss}$.  As indicated by the fluxon trajectories in the ($\upsilon$, 
$X$) phase plane, the $J^B_{th}/J_c^o$ versus $J^{sd}/J^{ss}$ curves for both 
$J^{sd}/J^{ss} > 0$ and $J^{sd}/J^{ss} < 0$ are identical.  This indicates that
the symmetry of order parameter (i.e., $S_{++}$ versus $S_{+-}$) may not be 
distinguished directly from the I-V curves.  The increase in the threshold bias
current with increasing interband Josephson current implies that, as the 
critical current modulation increases, a greater strength of driving force must
be provided by the bias current to allow the fluxon to pass through the region
of a static 2$\pi$-phase texture.

\section{summary and conclusion}

In summary, we have investigated the effects of interband Josephson current on
the fluxon dynamics of an LJJ with two-gap superconductors. Due to the 
multi-component nature of the superconducting order parameter, a 2$\pi$-phase 
texture may be present in each S layer.  Accounting for the charge imbalance 
between the two electronic bands, say $s$ and $d$ bands, we found that the 
dynamics of the relative phase of the $s$ and $d$ condensates may be described 
by the equation of motion which is similar to that for the fluxon.  However, the 
kink-solution for the relative phase (i.e., $i$-soliton) differs from that for 
the fluxon.  Unlike fluxon, an $i$-soliton does not carry a magnetic flux 
quantum but leads to both spatial and temporal modulation of the critical 
current density.    

The critical current density modulation, induced by excitation of an $i$-soliton
representing a 2$\pi$-phase texture, behaves as an effective pinning
potential for fluxon and modifies the fluxon trajectories.  These trajectories 
in the velocity-position phase plane are similar to those for fluxon moving in 
the presence of a single microresistor.  Moreover, similar to an array of 
microresistors in the insulator layer of a LJJ, a periodic modulation of the 
critical current may be created when a spatially periodic array of $i$-solitons
is excited.  This critical current modulation can change the fluxon speed when
the bias current is applied across the junction in a dissipative environment. 
In the region far away from the critical current modulation, the fluxon motion is uniform.  However, in the region of ciritical current modulation, the fluxon
speed can either decrease or increase from the uniform speed as the fluxon 
either receives or returns energy to the junction.  The decrease in the fluxon 
speed due to excitation of an $i$-soliton results in emission of EM radiation 
(i.e., quasi-linear plasma wave).  

The changes in the fluxon speed due to critical current modulation may be 
reflected in the I-V characteristics.  When the bias current is less than the 
critical value, the fluxon becomes trapped and its speed will be reduced to 
zero, resulting zero voltage across the junction.  This is similar to a fluxon 
trapped by a microresistor in the I layer.  The result suggests an interesting
possibility that a double-well potential for a fluxon, which is a necessary 
condition for a Josephson vortex quantum bit, may be created by exciting two 
$i$-solitons in each S layer of the LJJ, rather than implanting two closely 
spaced microresistors.\cite{KDP}  Also, the I-V curves reveal the dependence of
fluxon motion on the bias current.  The threshold bias current needed for the 
fluxon to overcome the critical current modulation appears as a discontinuous
jump in the I-V curve at V=0.  Hence, this discontinuity may serve as a way to 
verify the excitation of an $i$-soliton.  Since the size of this discontinuity
depends on the interband Josephson current, the I-V curves may be used to 
estimate the strength of the interband Josephson effect.  However, the I-V 
curves for both the $S_{++}$ and $S_{+-}$ symmetry superconductors are expected
to be similar since the threshold bias current does not depend on the symmetry
of the order parameter.  

The present work indicates that the changes in the fluxon dynamics due to 
excitation of an $i$-soliton in a single LJJ may become amplified in a multiple
LJJ stack where the collective motion of fluxon is known to arise.\cite{KP}
Furthermore, as recent studies of three-band superconductors indicate, 
excitation of $i$-solitons\cite{GCB} which break\cite{ST} the time reversal 
symmetry is possible when the interband Josephson couplings are 
frustrated.\cite{LH}  The fluxon dynamics of a LJJ with three band superconductors may lead to a more interesting junction property than that 
described in the present work.

\vskip 0.4in

This work was supported in part by National Science Foundation through North 
Dakota EPSCoR Grant No EPS-081442.  The authors thank W. Schwalm for helpful 
discussions. 
\vskip 0.3in

\centerline{ {\bf APPENDIX:}}
\vskip 0.1in

\centerline{DERIVATION OF GINSBERG-LANDAU ENERGY}
\centerline{FOR A TWO-GAP SUPERCONDUCTOR}
$\\$

For completeness, we derive the Ginsberg-Landau free energy from the two-gap 
Hamiltonian ${\hat {\cal H}}_{TB}$ of Eq. (\ref{two-gap}) by writing the 
partition function $\cal Z$ as
\begin{equation}
{\cal Z} = \int {\cal D}[{\bar c}^s, c^s] {\cal D}[{\bar c}^d, c^d] 
e^{-{\cal S}[{\bar c}^s, c^s,{\bar c}^d, c^d]}
\label{parti1}
\end{equation}
where the action $\cal S$ is given by
\begin{equation}
{\cal S} = \int_0^\beta d\tau \bigg[\sum_{i,{\bf k},\sigma} {\bar c}_{{\bf k},\sigma}^i 
\partial_\tau c_{{\bf k},\sigma}^i + {\hat {\cal H}}_{TB} \bigg]~,
\label{act1}
\end{equation}
$\partial_\tau = \partial/\partial\tau$ and ${\bar c}_{{\bf k}\sigma}^i$ 
($c_{{\bf k}\sigma}^i$) is the Grassmann variable representing the creation 
(annihilation) operator for the $i$-band electron in the (${\bf k} \sigma$) 
state.  Here, we suppress the S layer index $\ell$, for simplicity.  We 
proceed by introducing Nambu notation of ${\bar {\cal C}}^i_{\bf k}=
({\bar c}_{{\bf k}\uparrow}^i c_{-{\bf k} \downarrow}^i)$ and by writing the 
pair fields ${\bar{\cal A}}^i_{\bf k}$ and ${\cal A}^i_{\bf k}$ as
\begin{eqnarray}
{\bar{\cal A}}^i_{\bf k} &=& {\bar {\cal C}}^i_{\bf k} \tau_+ 
{\cal C}_{\bf k}^i =
{\bar c}_{{\bf k}\uparrow} {\bar c}_{-{\bf k} \downarrow} \\
{\cal A}^i_{\bf k} &=& {\bar {\cal C}}^i_{\bf k} \tau_- {\cal C}_{\bf k}^i =
{c}_{-{\bf k}\downarrow} {c}_{{\bf k} \uparrow}
\end{eqnarray}
where $\tau_\pm = (\tau_1 \pm i \tau_2)/2$ and $\tau_i$ are Pauli matrices.  
The action $\cal S$ of Eq. (\ref{act1}) may be rewritten as
\begin{equation}
{\cal S} = \int_0^\beta d\tau \sum_{i, {\bf k}} 
\bigg[ {\bar{\cal C}}^i_{\bf k}
 (\partial_\tau I + \epsilon_{\bf k}^a \tau_3 ) {\cal C}^i_{\bf k} + \sum_{j,{\bf k}'} 
V_{ij}{\bar {\cal A}}^i_{\bf k} {\cal A}^j_{{\bf k}'} \bigg] .
\end{equation}
We introduce the Hubbard-Stratonovich transformation to map the interacting 
system to non-interacting fermions moving in an Hubbard-Stratonovich field 
$\Phi$ (i.e., auxilary field) representing electron pairing.  We rewrite the 
partition function of Eq. (\ref{parti1}) as 
\begin{equation}
{\cal Z} = \int {\cal D}[{\bar {\cal C}}, {\cal C}] {\cal D}[{\bar \Phi}, \Phi]
e^{-{\cal S}[{\bar {\cal C}}, {\cal C},{\bar \Psi}, \Psi]}
\end{equation}
where the action $\cal S$ is given by
\begin{eqnarray}
{\cal S} = \int_0^\beta d\tau \bigg[ \sum_{i, {\bf k}} {\bar{\cal C}}^i_{\bf k}
 (\partial_\tau I + \epsilon_{\bf k}^i \tau_3 ) {\cal C}^i_{\bf k}~~~~~~~~~~~~~~~~ 
\nonumber \\
~~~~~~~~~~~~~~~+ \sum_{{\bf k},{\bf k}'} \bigg(
{\bar \Phi}_{\bf k} {1 \over V} {\Phi}_{{\bf k}'}
- {\bar {\cal A}}_{\bf k} {1 \over V} {\cal A}_{{\bf k}'} \bigg) \bigg] .
\end{eqnarray}
Here, we note that 
${\bar \Phi}_{\bf k}=({\bar \Phi}^s_{\bf k},{\bar \Phi}^d_{\bf k})$,
${\bar {\cal A}}_{\bf k}=({\bar {\cal A}}^s_{\bf k},{\bar {\cal A}}^d_{\bf k})$, and $V$ denotes the pairing interaction matrix
\begin{equation}
V = \left( \matrix{ V_{ss} & V_{sd} \cr
                    V_{sd} & V_{dd} \cr } \right)~.
\end{equation}
We shift the $\Phi$ field (i.e., 
${\bar \Phi}_{\bf k} \rightarrow {\bar\Phi}_{\bf k} + {\bar {\cal A}}_{\bf k} V$
and  ${\Phi}_{\bf k} \rightarrow {\Phi}_{\bf k} + V {\cal A}_{\bf k}$) and obtain
\begin{eqnarray}
{\cal S} = \int_0^\beta d\tau \bigg[ \sum_{i, {\bf k}} {\bar{\cal C}}^i_{\bf k}
 (\partial_\tau I + \epsilon_{\bf k}^i \tau_3 ) {\cal C}^i_{\bf k}~~~~~~~~~~~~~~~~ 
\nonumber \\
~~~~~~~+ \sum_{{\bf k},{\bf k}'} \bigg(
{\bar \Phi}_{\bf k} {1 \over V} {\Phi}_{{\bf k}'}
+ {\bar {\cal A}}_{\bf k}\Phi_{{\bf k}'} + 
{\bar \Phi}_{\bf k} {\cal A}_{{\bf k}'} \bigg) \bigg] .
\end{eqnarray}
We integrate out the fermion variables ${\bar{\cal C}}$ and $\cal C$ by using 
the Grassmann integrals and obtain
\begin{eqnarray}
{\cal S} = \int_0^\beta d\tau \sum_{{\bf k},{\bf k}'} \left[ \sum_i
{{\bar \Phi}^i_{\bf k} {\Phi}^i_{{\bf k}'} \over g_i}  -
{g_{sd} \over {g_{s}g_{d}}} ({\bar \Phi}^s_{\bf k} {\Phi}^d_{{\bf k}'}  +
{\bar \Phi}^d_{\bf k} {\Phi}^s_{{\bf k}'} )\right]
\nonumber \\
-{\rm Tr} \ln G_s^{-1} - {\rm Tr} \ln G_d^{-1}~~~~~~~~~~~~~~~~~~~~~
\label{act2}
\end{eqnarray}
where $g_i=(V_{ss}V_{dd} - V^2_{sd})/V_{ii}$ and $g_{sd}=V_{sd}(V_{ss}V_{dd} - V^2_{sd})/(V_{ss}V_{dd})$.  We note that the first term in Eq. (\ref{act2}) may
be considered as the pesudo-order parameter contribution ${\cal S}_{gap}$ to 
the action.  The interacting Green function $G_i$ for the $i$-band electron is
given by
\begin{equation}
G_i^{-1} = (G_i^o)^{-1} + \Sigma_i 
\end{equation}
where the non-interacting Green function $G_i^o$ is given by
\begin{equation}
(G_i^o)^{-1} =\left( \matrix{ \partial_\tau+\epsilon_{\bf k}^i & 0 \cr
                  0 & \partial_\tau-\epsilon_{\bf k}^i \cr } \right)~,
\end{equation}
and the pair interaction $\Sigma_i$ is given by
\begin{equation}
\Sigma_i =\left( \matrix{ 0 & \Phi^i_{\bf k} \cr
                  {\bar\Phi}_{\bf k}^i & 0 \cr } \right)~.
\end{equation} 
We may simplify Eq. (\ref{act2}) by expanding the terms ${\rm Tr} \ln G_i^{-1}$
as
\begin{equation}
\ln G_i^{-1} = \ln (G_i^o)^{-1} - \sum_{n=1}^\infty {(-1)^n \over n}
(G_i^o \Sigma_i )^n  .            
\end{equation}
In the expansion, all odd order terms vanish when the trace of $\ln G_i^{-1}$ 
is taken.  Assuming that the pair field $\Phi_{\bf k}^i$ is uniform, we may
write the fourth-order contribution in the expansion of $\ln G_i^{-1}$ as
\begin{eqnarray}
{1 \over \beta}{\rm Tr} ( G_i^o \Sigma_i )^4 &=& 
{1 \over \beta} \sum_\nu \int {d^3{\bf k} \over (2\pi)^3} 
{2 \vert \Phi^i \vert^4  \over [\omega_\nu^2 + (\epsilon_{\bf k}^i)^2]^2}
\nonumber \\
&=& {7\zeta(3) \rho_i \over 4\pi^2T^2} \vert \Phi^i \vert^4
\end{eqnarray}
by making the high density approximation.  Here, $\omega_\nu = 
(2\nu +1)\pi/\beta$ with $\nu = 0, \pm 1, \pm 2, \cdot\cdot\cdot$ denotes the 
Matsubara frequency, and $\zeta(x)$ is the zeta-function  The second order 
contribution to the expansion yields
\begin{equation}
{1 \over \beta}{\rm Tr}(G_i^o \Sigma_i)^2 = {2 \over \beta} 
\sum_{\nu, {\bf k}, {\bf k}'}
{\vert \Phi^i_{{\bf k} -{\bf k}'}\vert^2 \over 
(i\omega_\nu - \epsilon_{\bf k}^i)
(i\omega_\nu + \epsilon_{{\bf k}'}^i)}  .
\end{equation}
We may simplify the calculation by introducing new momentum variables: 
${\bf p} =({\bf k} +{\bf k'})/2$ and ${\bf q} = {\bf k} - {\bf k}'$.  We make a
Taylor expansion in $\bf q$ up to second order, in the high-density limit, and
obtain
\begin{eqnarray}
{1 \over \beta}{\rm Tr}(G_i^o \Sigma_i)^2 = {2 \over \beta} \sum_{\nu, {\bf q}} 
\vert \Phi^i_{\bf q}\vert^2 \sum_{\bf p}\bigg\lbrace
{ 1 \over \omega_\nu^2 - (\epsilon_{\bf p}^i)^2} ~~~~~
\nonumber \\
+ {{\bf p}^2{\bf q}^2 \over {12(m_i^o)^2}} 
\bigg[ {-3 \over (\omega_\nu^2 +(\epsilon_{\bf p}^i)^2)^2} +
{4(\epsilon_{\bf p}^i)^2 \over (\omega_\nu^2 + (\epsilon_{\bf p}^i)^2)^3} \bigg] \bigg\rbrace .
\label{secord}
\end{eqnarray}
Here, we used the dispersion relation of $\epsilon_{\bf k}^i ={\bf k}^2/2m_i^o$,
where $m_i^o$ is the band mass of the electron.  Also we used the relation 
$p_ip_j \rightarrow \delta_{ij} p^2/3$ when the angular average is performed.
In Eq. (\ref{secord}), the first term yields the quadratic term (i.e., 
$\vert \Phi^i \vert^2$) in the GL free energy, while the second term of leads 
to the gradient part of the quadratic term (i.e., $\vert\nabla\Phi^2 \vert^2$).
Combining these terms, we obtain
\begin{eqnarray}
{1 \over \beta}{\rm Tr}(G_i^o \Sigma_i)^2 =
 - 2\rho_i \ln {2\beta \Lambda e^\gamma \over \pi} \vert \Phi^i \vert^2 ~~~~~~~
\nonumber \\
~~~~~~~+ 
{7 \zeta(3) k_F^2 \beta^2 \rho_i \over 24 (m_i^o)^2 \pi^2} \int d^3{\bf r} 
\vert \nabla \Phi^i ({\bf r}) \vert^2
\end{eqnarray}
where $\Lambda$ is the cut-off energy for the boson which mediates pairing 
interaction and $\gamma \approx 0.5772$.  Here, we approximated that 
$p^2 \approx k_F^2$.  We now combine the result and write the coeffcients to 
the GL free energy expansion for ${\bar {\cal G}}^i_{OB}={\cal G}_{OB}^i -
{\cal G}_{ng}^i$ for one dimension in the superconducting state as
\begin{equation}
{\bar {\cal G}}^i_{OB}= a^{GL}_i \vert \Phi^i \vert^2 +
{b^{GL}_i \over 2}\vert \Phi^i \vert^4 + {1 \over 2m_i^*} \bigg\vert
{d\Phi^i \over dx}\bigg\vert^2 ~,
\label{GLOB}
\end{equation}
where the coefficients of the expansion are
\begin{eqnarray}
a_i^{GL} &=& {1 \over g_i} - \rho_i \ln {2\beta\Lambda e^{-\gamma} \over \pi} ,
\\ 
b_i^{GL} &=& {7\zeta(3)\beta^2\rho_i \over 16 \pi^2} ,
\end{eqnarray}
and the effective mass $m_i^*$ is
\begin{equation}
m_i^* = {24 (m_i^o)^2\pi^2 \over 7 \zeta(3) k_F^2 \rho_i}~.
\end{equation}
The GL free energy of Eq. (\ref{GLOB}) may be used to estimate the interband 
phase dynamics by noting that the Hubbard-Strotonovich field may be expressed 
in terms of modulus-phase variables as 
$\Phi_i= \vert \Phi^i \vert \exp (i\theta^i)$, as discussed in Sec. III.

\end{document}